\documentclass[twocolumn,preprintnumbers,amsmath,amssymb]{revtex4}

\usepackage{graphicx}
\usepackage{dcolumn}
\usepackage{bm}
\usepackage{subfigure}  

\IfFileExists{srcltx.sty}{\usepackage[active]{srcltx}}

\begin{document}


\title{Can past gamma-ray bursts explain both INTEGRAL and ATIC/PAMELA/Fermi anomalies simultaneously?
}

\author{Antoine  Calvez}
 \email{acalvez@ucla.edu}%
\affiliation{Department of Physics and Astronomy, University of California, Los
Angeles, CA 90095-1547, USA}
\author{Alexander  Kusenko}
\affiliation{Department of Physics and Astronomy, University of California, Los
Angeles, CA 90095-1547, USA}
\affiliation{IPMU,
University of Tokyo, Kashiwa, Chiba 277-8568, Japan}


\begin{abstract}
Gamma-ray bursts (GRBs) have been invoked to explain both the $511\,\mathrm{keV}$ emission from the galactic bulge and the
high-energy positron excess inferred from the ATIC, PAMELA, and Fermi data.
While independent explanations can be responsible for these phenomena, we explore the possibility of their common GRB-related origin by
modeling the GRB distribution and estimating the rates. For an expected Milky Way long GRB rate, neither of the two signals is generic;
the local excess requires a $2\%$ coincidence while the signal from the galactic center requires a $20\%$ coincidence with respect to the timing of the latest 
GRB. The simultaneous explanation 
requires a $0.4\%$ coincidence. Considering the large number of statistical ``trials'' created by  multiple searches for new physics, 
the coincidences of a few per cent cannot be dismissed as unlikely.  Alternatively, both phenomena can be explained by GRB if the galactic 
rate is higher than expected.  We also show that a similar result is difficult to obtain assuming a simplified short GRB distribution.

\end{abstract}

\pacs{98.70.Rz,95.85.Ry}
\maketitle

Gamma ray bursts are among the most energetic events in the universe.  
Although there is a significant correlation of long GRBs with star-forming metal-poor galaxies~\cite{Fruchter:2006py,Stanek:2006gc}, 
many long GRBs are observed in high-metallicity galaxies as well~\cite{Savaglio:2006xe,CastroTirado:2007tn,Levesque:2010rn}. 
Therefore, one expects that long GRBs should occur in 
the Milky Way Galaxy once every $10^{4}-10^{6}$ years~\cite{Schmidt:1999iw,Frail:2001qp,Furlanetto:2002sb,Perna:2003bi,Podsiadlowski:2004mt}.
This large uncertainty in the rate is largely due to selection effect, as the observed GRB rate and the opening solid angle necessary to determine the absolute rate
are sampled from two different populations. Furthermore, the large variations observed in the beaming angle, and the uncertainties in the measurement of those angles 
make it difficult to accurately constraint the absolute GRB rate in the Milky Way \cite{Podsiadlowski:2004mt,Frail:2001qp}.
Short GRBs, which probably result from compact star mergers, should occur in every galaxy, including the Milky Way,
at a rate that is comparable, although slightly lower. Short GRBs are likely less beamed then long GRBs, and some even appear to not be beamed at all
\cite{Grupe:2006uc,Burrows:2006ar,Watson:2006pf}. For the purpose of this work, we follow limits set using observation of $\gtrsim0.1\,\mathrm{mJy}$ sources 
at $8.44\,\mathrm{GHz}$ by the Very Large Array (VLA), and we assume a lower limit on the beaming angle of a few degrees \cite{Perna:1998ny}. 
This sets an approximate upper limit on the rate $R_{\rm GRB}\lesssim0.5\times10^{-4}$ per year. 
 
GRBs should produce high energy electrons and 
positrons~\cite{2000ApJ...534L.155D,Meszaros:2000cc,Furlanetto:2002sb}, 
and it is therefore natural to consider GRBs as a possible explanation of recent astro-particle observations involving 
electrons and positrons~\cite{Furlanetto:2002sb,Bertone:2004ek,Parizot:2004ph}, as well as cosmic rays~\cite{Calvez:2010uh}.

High energy electrons and positrons are produced during the initial burst \cite{1978MNRAS.183..359C,Shemi:1990rv}.  However, due to the short recombination time scale, these electrons are unlikely to escape the initial fireball \cite{Parizot:2004ph}. On the other hand, high energy electrons and positrons can also be produced by the interaction of the $\gamma$-rays with the surrounding medium in the jets \cite{Rees:2004gt,Ioka:2007qk}, or ahead of the fireball \cite{Thompson:1999vc,Meszaros:2000cc,2000ApJ...534L.155D}. In the former case the annihilation of the pairs freezes out as the jet expands, and electrons and positrons are allowed to escape to the outside. In the latter case $\gamma$-ray photons are back scattered
into the ejecta and allowed to pair produce through interactions with subsequent GRB photons. In both cases conversion efficiencies is expected to be on the order of $1-10\%$.

Results from PAMELA have shown an excess in the positron fraction above $10\,\mathrm{GeV}$ \cite{PAMELA},
while Fermi, PPB-BETS and ATIC have shown an excess in the total electron and positron flux above $100\,\textrm{GeV}$ \cite{Fermi,Torii:2008xu,ATIC}.
Both of these results can simultaneously be explained by a nearby ($\sim1\,\textrm{kpc}$) GRB-like event
$2\times10^{5}$ years ago \cite{Ioka:2008cv}. Most GRB models usually predict the electron positron pairs to form with $\mathrm{MeV}$ energies
\cite{2000ApJ...534L.155D,Thompson:1999vc}. However, the interaction of backscattered $\mathrm{TeV}$ photons created during the burst with optical photons from
the afterglow would also give rise to a $\mathrm{GeV}-\mathrm{TeV}$ population of electron and positrons. The total energy of the produced pairs would be
comparable to that of the  initial high energy photons: ~$10^{50}\,\mathrm{erg}$ for a typical spectral index of $\gamma=2.2$ \cite{Ioka:2008cv}.

Furthermore, observations by INTEGRAL satellite of the central region of the Galaxy have discovered a spectral line of energy $511\,\mathrm{keV}$
coming from the galactic  bulge~\cite{Knodlseder:2003sv,Jean:2003ci,Teegarden:2004ct,Churazov:2004as,Knodlseder:2005yq,Jean:2005af}.
This emission could also be explained by GRBs inside or near the Central Molecular Zone (CMZ) occurring every $\sim10^{4}$
years \cite{Parizot:2004ph,Bertone:2004ek}. Indeed, a conversion efficiency on the order of $1-5\%$, would generate a number of $\mathrm{MeV}$ positrons
\begin{equation}
N_{+}=\left(10^{54}-10^{55}\right)\left(\frac{E}{10^{51}\,\mathrm{erg}}\right),\nonumber
\end{equation}
where $E$ is the total initial energy of the burst
\cite{Thompson:1999vc,2000ApJ...534L.155D,Meszaros:2000cc,Bertone:2004ek,Parizot:2004ph}.  The electrons and positrons lose energy via ionization on a time scale of $\sim10^{7}$ years and diffuse through the bulge, traveling a distance similar to the extent of the $511\,\mathrm{keV}$ image before stopping~\cite{Bertone:2004ek,Parizot:2004ph}.  Thus, the explanation is in agreement with the morphology of the observed signal.  Since the probability of annihilation is much higher at low energies, most of the positrons injected into the central region slow down and annihilate at rest~\cite{Bertone:2004ek,Parizot:2004ph}. 

The explanations put forth to explain these two phenomena in Refs.~\cite{Bertone:2004ek,Parizot:2004ph} and in Ref.~\cite{Ioka:2008cv} have not made any contradictory assumptions. The morphology and the intensity of the $511\,\mathrm{keV}$ line depends on some (reasonable) assumptions regarding the gas density and magnetic fields in the Galactic Center~\cite{Bertone:2004ek,Parizot:2004ph}, which do not affect the high-energy signals~\cite{Ioka:2008cv}.  However, it has not been shown that the rates of GRBs and their distribution in the Galaxy afford simultaneous explanation of both the low-energy and the high-energy anomalies.  This is the question that we will address.

Gamma ray bursts form to distinct populations, long and short \cite{Kouveliotou:1993yx}. Although there remain a great deal of uncertainties regarding the origin of GRBs, long GRBs are usually associated with core-collapse supernovae in star forming galaxies \cite{Woosley:2006fn}, while short GRBs are usually attributed to the merger of neutron stars and/or black holes~\cite{Nakar:2007yr}.
We will explore the possibility that both type of GRBs may be responsible for the
$511\,\mathrm{keV}$ and the excess measured by ATIC, Pamela and Fermi, simultaneously.

\begin{figure}[ht!]
  \begin{center}
 \includegraphics[width=0.99 \columnwidth]{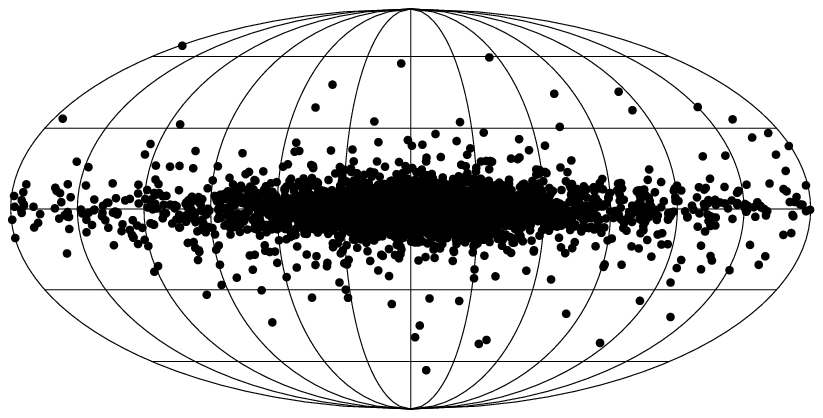}\\
\includegraphics[width=0.99 \columnwidth]{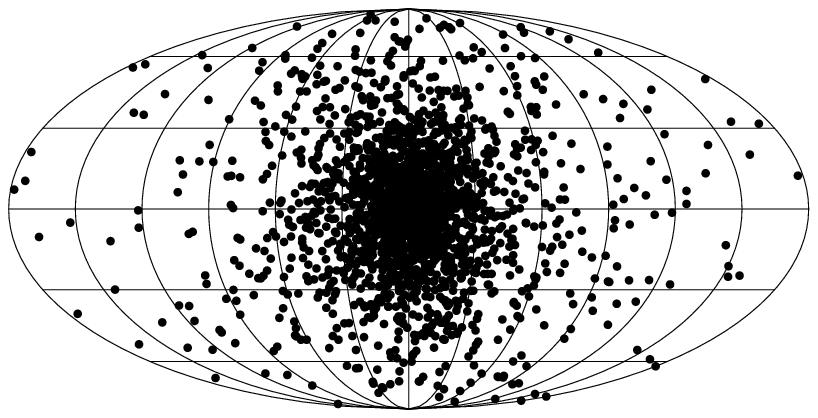}
\caption{Simulated map of GRB locations in galactic coordinates for a distribution that follows the star counts~\cite{Bahcall:1980fb,Bahcall:1985hi} (top), and for a distribution inferred
from observations of short GRBs~\cite{Cui:2010pv} (bottom)\label{fig:skymap}.
  }
    \end{center}
\end{figure}

We model the history of Galactic GRBs in a Monte Carlo simulation by randomly generating bursts to determine the probability that both the electron-positron excess in the ISM, and the $511\,\textrm{keV}$ line might originate from GRBs.
To this end, we calculate the probability that a measurement made at a random time in the history of our galaxy falls within the required time constraints.  We will assume that the spatial distribution of long GRBs follows the distribution of stars determined from the star counts~\cite{Bahcall:1980fb,Bahcall:1985hi}.  For the short GRBs, which occur farther in the halo, we will use the distribution of sources obtained from observations~\cite{Cui:2010pv}.  We show the simulated maps of the GRBs locations in FIG. \ref{fig:skymap}.

It is reasonable to assume that the distribution of long GRBs in the galaxy is closely related to the stellar density profile.
We model the stellar distribution in the galaxy as a two-components system comprised of a thin disk and a bulge.
Below we will also include the effects of the spiral arms.
We neglect the spheroid component because it  accounts for only $1-5\%$ of the mass of the galaxy.
We use the stellar density profiles of the Bahcall--Soneira model~\cite{Bahcall:1980fb,Bahcall:1985hi}:
\begin{eqnarray}
n_{D}&=&\rho_{D}(r_{0})\int drdzd\varphi\int dM\, \Phi(M) \times \nonumber \\ & &  r e^{-z/H(M)} e^{-(r-r_{0})/h},\\
n_{B}&=&\rho_{B}\int dRd\theta d\varphi\int dM\, \Phi(M)R^{0.2}\sin\theta e^{-R^{3}},
\end{eqnarray}
Where $z$ is the perpendicular distance from the plane of the disk, $r$ is the galactocentric distance with $r_{0}=8\,\textrm{kpc}$
the galactocentric distance of the sun, and $R$ is the spherical distance from the center of the galaxy.
Following the literature we take $h=3.5\,\textrm{kpc}$~\cite{Bahcall:1980fb,Bahcall:1985hi}.
The normalization constants $\rho_{D}$ and $\rho_{B}$ are only relevant in our work for determining the relative
fraction of stars in each components of the Milky Way. Based on mass studies, we assume that $85\%$ of the stars are located in the
disk and the remainder in the bulge. The luminosity function is given by
\begin{equation}
\Phi(M) = \left\{
\begin{array}{ll}
\frac{n_{*}10^{\beta\left(M-M^{*}\right)}}{\left[1+10^{\delta\left(\alpha-\beta\right)\left(M-M^{*}\right)}\right]^{1/\delta}}, & -6 \leq M \leq 15\nonumber\\
{\rm const}=\Phi(15), &  15 \leq M \leq 19\nonumber\\
0, & M < -6 \text{ or } M > 19\nonumber
\end{array} \right.
\end{equation}
where $n_{*}=4.03\times10^{-3}$, $M^{*}=1.28$, $\alpha=0.74$, $\beta=0.04$, and $\delta=1/3.4$.
Finally,  to facilitate the random sampling process, we adopt a simplified version of the scale height.
Old stars with visual magnitude $M>5.1$ are generally associated with an exponential scale height of $325\,\textrm{pc}$.
Main sequence stars with visual magnitude $M<2.3$, are well described by an exponential scale height of $H=90\,\textrm{pc}$;
for $2.3<M<5.1$, the scale height is interpolated with a linear regressions \cite{Bahcall:1980fb,Bahcall:1985hi}.
Instead of performing this regression we will, for the purpose of this work, assume the scale height to be well described by a step function.
In other words:

\begin{equation}
H(M) = \left\{
\begin{array}{rl}
90\,\textrm{pc} & M \leq 3.7\nonumber\\
325\,\textrm{pc} & M > 3.7.\nonumber
\end{array} \right.
\end{equation}

We will model the distribution of short GRBs based on a simplified star formation disk model \cite{Bloom:2000pq}. This assumption, although
not completely accurate, reflects the fact that a significant fraction of short GRBs could originate from massive star
collapse \cite{Virgili:2009ca,Zhang:2009uf}.

\begin{equation}
n_{\text{Sh}}=\rho_{\text{Sh}}\int dRd\theta d\varphi\, R e^{-1.67\beta(R)},
\end{equation}
where $\beta(R)=\frac{R}{R_{\text{half}}}$, with $R_{\text{half}}$ the half-light radius of the Milky Way.
We treat each coordinate as completely independent and, after normalizing the stellar density profiles such that the area
under the curves is equal to one, we treat the distributions as probability density functions.
For the purpose of the Monte Carlo we distribute the bursts over time using a Gaussian profile with average rate
\begin{equation}
R_{_{\rm GRB}} = r_{_{4,\rm GRB}} \ \frac{1}{10^{4} {\rm yr}}= (0.1-0.5) \ \frac{1}{10^{4} {\rm yr}}.
\end{equation}

We first proceed with the long GRB analysis. We randomly generated $10^{6}$ bursts based on the spatial and temporal distribution described above,
and we recorded the number of events (i) occurring within $1\,\textrm{kpc}$ of the sun, as well as those (ii) inside CMZ or beamed towards the CMZ.
We find that about $0.2\%$ of all events fall in category (i). Similarly, we find that about $12\%$ of all even fall in category (ii),
and hence, the average rate of GRB, and, therefore, of electrons and positrons injection in the central part of the bulge is
\begin{equation}
R_{_{\rm CMZ}}=\frac{1}{8.5\times 10^4 {\rm yr}} \left( \frac{r_{_{4,\rm GRB}} }{1.0}\right)\label{R_CMZ}\\
\end{equation}

Let us now discuss the probability of a GRB within $1\,\textrm{kpc}$ of the sun.
For a uniform distribution of stars in the disk, we obtain the rate of nearby GRB
 \begin{equation}
R_{\odot,{\rm \text{uniform}}}=\frac{1}{5\times 10^6 {\rm yr}} \left( \frac{r_{_{4,\rm GRB}} }{1.0}\right)
\end{equation}

However, the sun is located in a spiral arm, and the local density of stars is expected to be somewhat higher.
Following Ref.~\cite{ortiz:1993A&A}, we model the Milky Way spiral arms of the form:
\begin{equation}
r(\theta)=q_{0}e^{(\theta-\theta_{0})\tan i}, \nonumber
\end{equation}
where $q_{0}=2.3\,\textrm{kpc}$, $i=13.8^\circ$, and $\theta_{0}$ ranging from $0$ to $\frac{3\pi}{2}$ in steps of $\frac{\pi}{2}$
representing the starting angle of each arm. The brightness of the arms varies with the passband it is observed in. In the $\rm O$ passband
the contribution of the arms to the total brightness increases from $17\%$ at $3\,\mathrm{kpc}$ to $50\%$ at $15\,\mathrm{kpc}$; those contributions are
$20\%$ stronger in the $\rm B$ passband, and $50\%$ stronger in the $\rm A$ passband \cite{BinneyGA}. Given the location of the solar system, we find that this 
over-density results in a factor $f_{\rm spiral }\sim 2$ enhancement in the local GRB rate:
 \begin{equation}
R_{\odot} \sim 2\, \left(\frac{f_{\rm spiral}}{2}\right) R_{\odot,{\rm \text{uniform}}}\sim \frac{1}{2.5\times 10^6 {\rm yr}} \left( \frac{r_{_{4,\rm GRB}} }{1.0}\right).
\label{R_solar}
\end{equation}

Let us now compare the rates in Eqs.~(\ref{R_CMZ}) and (\ref{R_solar}) with the mode requirements.
As discussed in Ref.~\cite{Parizot:2004ph}, the positron population in the bulge enters a steady-state regime at the right density to explain
the $511\,\textrm{keV}$ for $R_{_{\rm CMZ}} \gtrsim (3\times10^{4} {\rm yr})^{-1}$.  For a lower rate, the positron population rises and dies off after each GRB.
Matching this rate to Eq.~(\ref{R_CMZ}) requires $r_{_{4,\rm GRB}}\approx 3$, which outside the
expected range $r_{_{4,\rm GRB}}=0.1-0.5 $ \cite{Frail:2001qp}.

Of course, there is no reason to believe that the rate of $511\,\textrm{keV}$ photons is time-independent and that we don't just happen to observe it at the peak
shortly after a GRB. In this case, one can estimate the probability of such a coincidence. If GRBs are responsible for the $511\,\textrm{keV}$ line, and if one
takes $r_{_{4,\rm GRB}}=0.5 $, than one requires that the current observation be within the last $20\%$ of the average GRB time separation,
which is about a $20\%$ coincidence.

For a local GRB to explain the high-energy electrons and positrons, a GRB should have happened in the solar neighborhood
about $\sim 2\times 10^5 {\rm yr}$ ago~\cite{Ioka:2008cv}. To match the rate in Eq.~(\ref{R_solar}), one must assume
$r_{_{4,\rm GRB}}\approx 12$, which is, again, outside the expected range $r_{_{4,\rm GRB}}=0.1-0.5$~\cite{Frail:2001qp}.
We estimate that a $\sim 2\%$ coincidence is required for a sufficiently recent GRB to explain ATIC PAMELA and Fermi data.

Finally, we ask what would be required to explain both observations by GRB. If a GRB occur at a much higher rate,
$R_{_{\rm GRB}}=12/10^4 {\rm yr}$, then both observations have a natural explanation by GRBs. If the GRB rate is lower, as expected, namely 
$R_{_{\rm GRB}}=(0.1-0.5)/10^4 {\rm yr}$, then a $2\%$ coincidence is required to observe a
local GRB sufficiently recently while the same requirement applied to a cenral Galactic GRB requires a $20\%$ coincidence. Of course, a simultaneous explanation requires a $\sim0.4\%$ coincidence in this case. 
This probability is by no means small. Many independent observations are carried out in search of new physics, including dark matter. 
The large number of independent searches increases the ``multiple trials`` effect, so that one should expect $0.4\%$ effects to be promoted by multiple trials.  Finally, it is possible that the rate 
mismatch is ameliorated by the model uncertainties discussed in Refs.~\cite{Bertone:2004ek,Parizot:2004ph,Ioka:2008cv}.  

We repeated the previous procedure for short GRBs, ignoring the overdensity effects of the arms, and found:
\begin{eqnarray}
R_{_{\rm CMZ}}^{\text{sh}}&=&\frac{1}{4.5\times 10^4 {\rm yr}} \left( \frac{r_{_{4,\rm GRB}} }{1.0}\right)\label{R_sh}\\
R_{\odot}^{\text{sh}}&=&\frac{1}{1.5\times 10^8 {\rm yr}} \left( \frac{r_{_{4,\rm GRB}} }{1.0}\right)
\end{eqnarray}

Therefore, in order to explain the $511\,\textrm{keV}$ we need $r_{_{4,\rm GRB}}\approx 1.5$ or a $35\%$ coincidence assuming $r_{_{4,\rm GRB}}=0.5 $, while
in order to explain the electron and positron excess we need $r_{_{4,\rm GRB}}\approx 750$, or a $0.067\%$ coincidence. In other words, the simultaneous explanation would require in this case a $0.023\%$ coincidence. This results implies that the short GRBs are less likely to account for both effects.

We conclude that a simultaneous explanation of the $511\,\textrm{keV}$ line on one hand, and ATIC, PAMELA and Fermi on the other hand, based on GRBs in the Milky Way Galaxy is possible in one of two ways.  First, it is possible if the long GRB rate in our Galaxy is $3/10^4 {\rm yr}$, which is higher than expected. Second, it is possible if we are observing the effects of two very recent GRB, one in the Galactic Center and one in the local neighborhood, each of which happened more recently than average. The probability of the latter is about $\sim2\%$. It is however unlikely that short bursts be simultaneously responsible for both effects. Although short GRBs could easily explain the $511\,\textrm{keV}$ line, it is hard to reconcile them being the cause of the local electron and positron excess.

This work was supported by DOE Grant DE-FG03-91ER40662 and NASA ATP Grant  NNX08AL48G.  A.K. thanks Aspen Center for Physics for hospitality.

\bibliography{GRB}

\end{document}